\documentclass[fleqn,twoside]{article}
\usepackage{espcrc2}

\usepackage{amssymb}

\newcommand{\nat}{Nature}
\newcommand{\apjs}{ApJS}
\newcommand{\apj}{ApJ}
\newcommand{\apjl}{ApJ}
\newcommand{\aap}{A\&A}
\newcommand{\mnras}{MNRAS}
\newcommand{\apss}{ApSS}

\usepackage{graphicx}
\title{Thermonuclear X-ray Bursts: Theory vs.~Observations}

\author{Andrew Cumming\address{Hubble Fellow, Department of
Astronomy and Astrophysics, University of California, Santa Cruz;
cumming@ucolick.org}}

\begin{document}

\begin{abstract}
I review our theoretical understanding of thermonuclear flashes on
accreting neutron stars, concentrating on comparisons to
observations. Sequences of regular Type I X-ray bursts from GS~1826-24 and
4U~1820-30 are very well described by the theory. I discuss recent
work which attempts to use the observed burst properties in these
sources to constrain the composition of the accreted material. For
GS~1826-24, variations in $\alpha$ with accretion rate indicate that
the accreted material has solar metallicity; for 4U~1820-30, future
observations should constrain the hydrogen fraction, testing
evolutionary models. I briefly discuss the global bursting behavior of
burst sources, which continues to be a major puzzle. Finally, I turn
to superbursts, which naturally fit into the picture as unstable
carbon ignition in a thick layer of heavy elements. I present new
time-dependent models of the cooling tails of superbursts, and discuss
the various interactions between superbursts and normal Type I bursts,
and what can be learned from them.
\vspace{1pc}
\end{abstract}

\maketitle

\section{INTRODUCTION}

Type I X-ray bursts were discovered in the 1970's, and quickly
understood as being due to unstable burning of hydrogen and helium on the
surface of an accreting neutron star (for reviews see
\cite{Lewin95,StrohBild03}). Fuel accumulates for hours to days, and
then ignites and burns in a $\approx 10$--$100\ {\rm s}$ burst with
typical energy $\approx 10^{39}$--$10^{40}\ {\rm ergs}$.

In recent years, there has been a lot of excitement in the
field. First, the Rossi X-Ray Timing Explorer (RXTE) discovered
nearly-coherent oscillations during bursts, with frequencies in the
range $270$--$620\ {\rm Hz}$ \cite{StrohBild03}. Believed to be
rotationally-modulated brightness asymmetries on the neutron star
surface, the oscillation frequency tells us the neutron star spin, and
potentially about the spreading of burning, and the surface magnetic
field \cite{Chakrabarty03}.

Second, long term monitoring of bursters with BeppoSAX/WFC (Verbunt,
this volume) and RXTE/ASM and PCA has revealed a new class of
rare, extremely energetic, and long duration bursts now known as
``superbursts'' (\cite{Kuulkers02}; Kuulkers, this volume), as well as
a new class of low luminosity burst sources discovered through their
burst rather than persistent emission (\cite{Cornelisse02};
Cornelisse, this volume). Particularly important has been the large
amount of data accumulated on bursters by the Wide Field Camera (WFC)
survey of the Galactic center, which has also enlarged our sample of
normal Type I bursts \cite{Cornelisse03}.

In this article, I review our theoretical understanding of
thermonuclear flashes on accreting neutron stars, concentrating on how
well it compares to observations, and what we can learn from such a
comparison.

\section{HOW THE ACCRETED HYDROGEN AND HELIUM BURNS}
\label{sec:accum}
\label{sec:expected}

The physics of nuclear burning on the surface of an accreting neutron
star has been discussed by many authors, comprehensive reviews can be
found in refs.~\cite{Lewin95,Bildsten98,StrohBild03}. Here I briefly
review how the accreted hydrogen and helium burns, and the expected
burst properties as a function of accretion rate. In particular, I
concentrate on the differences between hydrogen and helium burning,
which are crucial for understanding the regimes of burning, and for
comparing to observations.

{\bf The timescale for burning} --- whereas helium burning proceeds
rapidly via strong interactions, hydrogen burning is relatively slow,
since beta decays are required to convert protons to neutrons. This
leads to a saturation of the CNO burning rate at high temperatures
$T\gtrsim 8\times 10^7\ {\rm K}$ (the so-called ``hot CNO cycle'';
\cite{HoyleFowler65}), when proton captures in the cycle occur more
rapidly than the subsequent beta-decays. At high accretion rates, hot
CNO burning during accumulation of fuel depletes hydrogen prior to the
runaway. The time to burn all the hydrogen in a given fluid element is
\begin{equation}\label{eq:tH}
t_H=11\ {\rm hrs}\ \left({0.02\over Z}\right)\ \left({X_0\over 0.7}\right),
\end{equation}
where $Z$ is the CNO mass fraction. 

The amount of hydrogen vs.~helium present at ignition of the burst has
a direct impact on the burst lightcurve. If the helium fraction is
large, the rapid energy release at the beginning of
the burst creates a luminosity exceeding the Eddington luminosity,
driving outwards expansion of the photosphere (``photospheric radius
expansion'') \cite{Fujimoto87}. If hydrogen dominates the composition however, it burns
after ignition via the rp-process
\cite{WallaceWoosley81,Schatz98}. Starting with seed nuclei made by helium burning, this comprises a series of proton captures and
beta-decays involving nuclei close to the proton drip line and
extending well beyond the iron group. This can significantly delay the
cooling of the burning layer, giving an extended burst tail, and
burst durations $\sim 100\ {\rm s}$ \cite{HanawaFujimoto84,Woos03}.

{\bf Thermal stability} --- The nature of the instability is a thin
shell flash \cite{HvH75}, in which temperature perturbations drive
runaway heating through temperature sensitive nuclear reactions. Hot
CNO hydrogen burning is {\it thermally stable}, since the burning rate
does not respond to temperature fluctuations. However, at the lowest
accretion rates $\dot M\lesssim 0.01\ \dot M_{\rm
  Edd}$\footnote{depending on metallicity, see \cite{Bildsten98} for
  the scalings. I take $\dot M_{\rm Edd}=1.7\times 10^{-8}\ M_\odot\
  {\rm yr^{-1}}$, the Eddington accretion rate for a 10 km neutron
  star accreting solar composition material.}, $T<8\times 10^7\ {\rm
  K}$ in the accumulating layer and the usual CNO cycle operates,
allowing unstable hydrogen ignition and leading to
``hydrogen-triggered'' bursts. For $\dot M\gtrsim 0.01\ \dot M_{\rm
  Edd}$, the hydrogen burns stably, and helium burning at a density
$\gtrsim 10^5\ {\rm g\ cm^{-3}}$ drives the thermal instability.

Helium triggered bursts are of two types, depending on the time to
accumulate the critical column of fuel (or equivalently the recurrence
time of the bursts $t_{\rm recur}$). If $t_{\rm recur}>t_H$, hydrogen
burns away and ignition occurs in a pure helium layer below a steady
burning hydrogen shell, giving a ``pure He'' burst; if $t_{\rm
  recur}<t_H$, hydrogen is present throughout the layer when the
helium ignites, giving a ``mixed H/He'' burst.

For $\dot M\gtrsim\dot M_{\rm Edd}$, helium burns at a temperature
$T\gtrsim 5\times 10^8\ {\rm K}$ for which the temperature sensitivity
of the triple alpha reaction is less than the temperature sensitivity of
the cooling. The burning is then thermally stable, and the fuel burns
at the rate it accretes. The accretion rate at which burning
stabilizes and its dependence on metallicity is still not
well-determined theoretically.

{\bf Nuclear energy release} --- hydrogen burning releases
substantially more energy per gram, $Q_{\rm nuc}\approx 7$ MeV per
nucleon compared to only $\approx 1.6$ MeV per nucleon from helium
burning. Observationally, the energy release per gram is measured by
the parameter $\alpha$, the ratio of persistent fluence between bursts
to burst fluence,
\begin{equation}\label{ratio}
\alpha\equiv\frac{\int_0^{\Delta t} F_{\rm p}\,dt}{\int_0^{\Delta t} F_{\rm
      b}\,dt} \approx {(GM/R)\over Q_{\rm nuc}},
\end{equation}
where $F_{\rm p}$ is the persistent flux from accretion, $F_{\rm b}$
is the burst flux, $\Delta t$ is the interval from the beginning of
one burst to the next, and $GM/R\approx 200$ MeV per nucleon is the
gravitational energy release. We expect $\alpha\approx 40$ if the
material burned in the burst is solar composition, compared to
$\alpha\gtrsim 100$ for pure helium. (These numbers are somewhat
uncertain given that not all of the accretion energy necessarily
appears in X-rays, and the X-ray emission is potentially
anisotropic). Measurement of $\alpha$ values in the expected range
was an important initial confirmation of the thermonuclear flash
model for Type I bursts.

\begin{figure}[htb]
%\vspace{9pt}
%\framebox[55mm]{\rule[-21mm]{0mm}{43mm}}
%\includegraphics[scale=0.6]{hotcnob}
\includegraphics[scale=0.44]{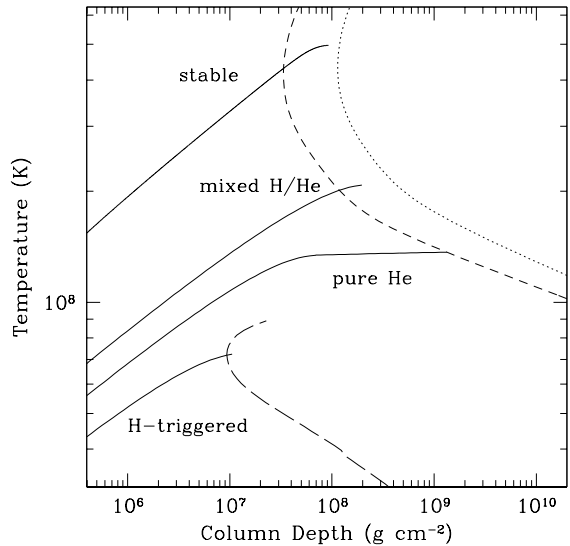}
\vspace{-1 cm}
\caption{Illustrative temperature profiles (solid curves) in the different burning regimes, for solar metallicity. The dotted and short-dashed lines are helium ignition curves (helium mass
fraction $Y=0.3$ and $Y=1.0$ respectively); the long-dashed line is the hydrogen ignition curve.}
\label{fig:regimes}
\end{figure}

The large energy release from hydrogen burning plays an important role
in setting the conditions for ignition of the flash. Figure
\ref{fig:regimes} shows illustrative temperature profiles as a
function of column depth (${\rm g\ cm^{-2}}$) into the star at the
moment of ignition for each of the four different burning regimes for
the case of solar metallicity. The dashed and dotted curves are
ignition curves for hydrogen and helium, calculated using a local
approximation for the cooling rate following ref.~\cite{CB00} (I refer
to that paper for details, also see \cite{NarayanHeyl03} for recent
calculations of the global eigenfunctions).

In the mixed H/He burst regime for solar metallicity, the ignition
conditions are well-determined, since the temperature profile is set
by hot CNO burning within the accumulating layer. The ignition column
depth is $\approx (1$--$2)\times 10^8\ {\rm g\ cm^{-2}}$, roughly independent
of accretion rate, so that bursts in this regime are expected to have
$\approx$ constant energy, and a recurrence time which scales close to
$\propto 1/\dot M$.

For low metallicity mixed H/He bursts, pure He bursts, and
hydrogen-triggered bursts, the internal heating from hot CNO burning
is either much less or absent altogether. The temperature profile is
then set by compressional heating or heating from below, either heat
from reactions occurring deep in the neutron star crust
\cite{Brown00}, or residual heat from or burning of leftover fuel in
the ashes of a previous burst \cite{Taam80,Woos03}. The ignition
conditions are therefore sensitive to accretion rate, and might be
expected to show variations from burst to burst (i.e.~less regular
bursting). This is particularly true for pure He bursts because of the
very shallow slope of the ignition temperature with column depth.

\section{COMPARISON WITH OBSERVATIONS}

\subsection{Regular bursting I: GS 1826-24, The ``Clocked Burster''}\label{sec:1826}

The transient source GS 1826-24 was discovered by Ginga, and regularly
monitored by BeppoSAX, which revealed it to be an extremely regular
Type I burster. Ubertini et al.~\cite{Ubertini99}, who dubbed this source
the ``clocked'' burster, found the burst recurrence time was close to
$6$ hours with a dipersion of only $\approx 6$ minutes. Further
monitoring showed that the recurrence time decreased as the source
brightened \cite{Cornelisse03}.

\begin{figure}[htb]
%\vspace{9pt}
%\framebox[55mm]{\rule[-21mm]{0mm}{43mm}}
%\includegraphics[scale=0.6]{hotcnob.ps}
\includegraphics[scale=0.46]{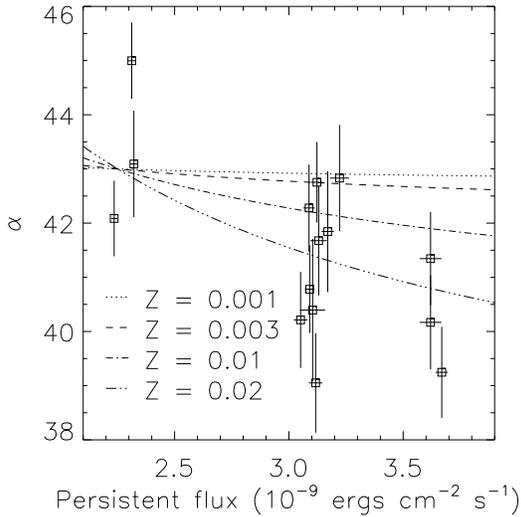}
\vspace{-0.8 cm}
\caption{Ratio of persistent to burst luminosity $\alpha=L_{\rm
p}/L_{\rm b}$ (equation \ref{ratio}), calculated from RXTE
observations between 1997 and 2002. Error bars represent the estimated
$1\sigma$ uncertainties. From Galloway et al.~(2003)
(ref.\cite{Gall03}). The changing $\alpha$ with $\dot M$ indicates
that fuel is burning between bursts, as expected for hot CNO hydrogen
burning with solar metallicity.}
\label{fig:gs1826alpha}
\end{figure}

Based on the energetics and recurrence times, as well as the long
duration of the bursts, Bildsten \cite{Bildsten00} proposed that this
was a textbook case of mixed H/He bursting. This is the expected
burning regime at the inferred accretion rate of $0.1\ M_{\rm Edd}$
and observed recurrence times of $<6$ hours, and is consistent with
the measured $\alpha=40$--$60$. In addition, the inferred ignition
mass is as expected from theory. The burst energy of $\approx 5\times
10^{39}\ {\rm erg}$ implies an ignition mass of $\approx 10^{21}\
{\rm g}$ for a nuclear energy release $5\times 10^{18}\ {\rm erg\
  g^{-1}}$ (equivalently, this is the mass accreted in a recurrence
time at the observed $\dot M$). The corresponding column depth is
$y_{\rm ign}\approx 10^8\ {\rm g\ cm^{-2}}$, exactly as
expected for mixed H/He ignition with solar metallicity ($Z=0.02$)
(Figure \ref{fig:regimes}; see also Table 2 of \cite{CB00}).

%\begin{figure}[htb]
%\vspace{9pt}
%\framebox[55mm]{\rule[-21mm]{0mm}{43mm}}
%\includegraphics[scale=0.6]{hotcnob.ps}
%\includegraphics[scale=0.43]{kepler1.ps}
%\vspace{-1.1 cm}
%\caption{Light curves for mixed hydrogen/helium burst for different
%choices of the nuclear physics (from ref.~\cite{Woos03}).}
%\label{fig:kepler1}
%\end{figure}

\begin{figure}[htb]
%\vspace{9pt}
%\framebox[55mm]{\rule[-21mm]{0mm}{43mm}}
%\includegraphics[scale=0.6]{hotcnob.ps}
\includegraphics[scale=0.4]{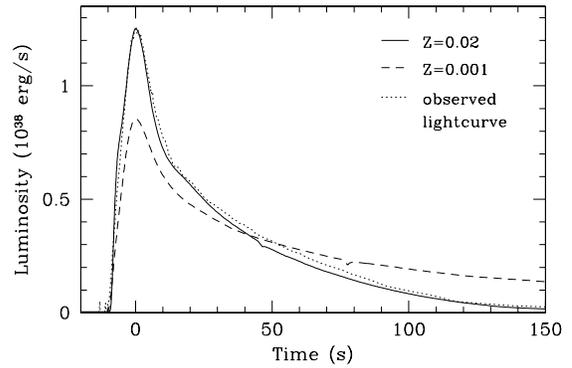}
\vspace{-1.1 cm}
\caption{The average observed lightcurve from GS~1826-24 (dotted line; ref.~\cite{Gall03}) compared with theoretical simulations (solid and dashed lines; models ZM and zM of ref.~\cite{Woos03}). The observed and theoretical recurrence times are both $\approx 4$ hours.}
% The models have $\dot M=0.1\ \dot M_{\rm Edd}$, and I adopt a gravitational redshift $z=0.26$.}
\label{fig:kepler1}
\end{figure}

Given the excellent agreement with theory, a natural question is
whether we can use the bursts to constrain properties of the source
\cite{Gall03}. For mixed H/He bursts, the ignition depth is mainly
sensitive to the metallicity, because the temperature profile of the
accumulating layer is set by hot CNO burning. This suggests an
argument to constrain the metallicity: if it was less than solar, the
reduced CNO heating would give a larger ignition column depth, and a
larger recurrence time and burst energy than observed. However, the
uncertainty in the relation between $\dot M$ and the X-ray luminosity
$L_X$ allows lower metallicity models to be accommodated by adopting a
larger $\dot M$ (the larger column is then accreted in the observed
recurrence time)\footnote{In fact, the best approach is to turn this
  around: given a metallicity, we ``measure'' the local accretion rate
  onto the star by matching the observed recurrence
  time.}. Additionally, the time-dependent simulations of Woosley et
al.\cite{Woos03} show that at low metallicity, a new heat source takes
over. Production of CNO elements in the tail of the previous X-ray
burst leads to leftover hydrogen burning beneath the newly accreted
layer (``compositional inertia''), reducing the sensitivity to
metallicity (see also \cite{HanawaFujimoto84}).

Determining the metallicity requires an additional constraint on the
models. Galloway et al.~\cite{Gall03} recently concluded that the
metallicity in the accreted layer was solar by looking at {\it
  variations of burst properties with $\dot M$}. They present data for
24 bursts observed by RXTE between 1997 and 2002.  The $\alpha$ value
was found to decrease by $\approx 10$\% as accretion rate increased by
$\approx 50$\% (Figure \ref{fig:gs1826alpha}). This is exactly as
expected from models with solar metallicity \cite{CB00}, since as
accretion rate increases and recurrence time drops, less time is
available to burn hydrogen between bursts. Models with metallicity
much less than solar are hard to reconcile with the $\alpha$
variations.

Another constraint comes from the burst lightcurves, which show a long
$\approx 10\ {\rm s}$ rise and a long $\approx 100\ {\rm s}$
tail\footnote{The burst lightcurves were remarkably similar during
  this period, with only a small variation as $\dot M$ increased with
  time. This, together with the extremely regular nature of the
  burning, argues for complete burning of the fuel in each burst, and
  complete covering of the stellar surface.}. Theoretical models of mixed H/He bursts (e.g.~\cite{HanawaFujimoto84,Schatz01}) show long burst tails, leading Bildsten
\cite{Bildsten98} to argue that an active rp-process powers the bursts from GS~1826-24. 
Figure \ref{fig:kepler1} compares the average observed burst lightcurve with the recent simulations of Woosley et al.~\cite{Woos03}. These simulations are the first to incorporate a large nuclear reaction network that can follow the detailed rp process nucleosynthesis together with resolved vertical structure. The agreement between the observed lightcurve and solar metallicity model is remarkable, supporting the findings of Galloway et al.~\cite{Gall03}. A more detailed comparison with the simulations is in progress. The predicted burst lightcurves are sensitive to the nuclear physics input \cite{Woos03}, and so there is potentially much to learn.

The ignition column for mixed H/He bursts is predicted to be roughly
independent of accretion rate, since hot CNO burning sets the
temperature in the accreted layer (\S 2).  Galloway et
al.~\cite{Gall03} find that $t_{\rm recur}\propto \dot M^{-1.05\pm
  0.02}$, implying a slight decrease in the ignition column with $\dot
M$, $y_{\rm ign}\propto \dot M^{-0.05}$. This remains to be
understood; a simple argument suggests a slight {\it increase} should
be seen, since less helium is produced by hot CNO burning for a
shorter recurrence time, requiring a slightly thicker layer to
ignite. Two possible explanations for the discrepancy are (i) extra heating, e.g.~from thermal or compositional inertia
effects \cite{Taam80,Woos03}, or (ii) the fraction of the star covered
by fuel changes with $\dot M$. Future comparisons with time-dependent
simulations or improved spectral models should test these ideas.

\subsection{Regular bursting II: 4U 1820-30}\label{sec:1820}

The ultracompact binary 4U~1820-30 (orbital period 11.4 mins
\cite{SPW87}) is an interesting source in which to study Type I
bursting behaviour.  It has a known distance, being located in the
metal-rich globular cluster NGC~6624 ($[$Fe$/$H$]\approx -0.4$,
distance $7.6\pm 0.4\ {\rm kpc}$) \cite{Kuulkers03}. It undergoes a
characteristic 176 day accretion cycle \cite{Pried84} during which the
accretion rate varies by a factor of 3 or more. Regular sequences of
Type I bursts are observed during the low state, with recurrence times
of $\approx 2$--$4$ hours
\cite{Clark76,Clark77,Haberl87,Cornelisse03}. In the high state,
bursts become irregular or disappear
altogether\cite{ChouGrindlay01,Cornelisse03}. This source also showed
the most energetic and luminous superburst \cite{StrohBrown}.

\begin{figure}[htb]
%\vspace{9pt}
%\framebox[55mm]{\rule[-21mm]{0mm}{43mm}}
%\includegraphics[scale=0.6]{hotcnob.ps}
\includegraphics[scale=0.5]{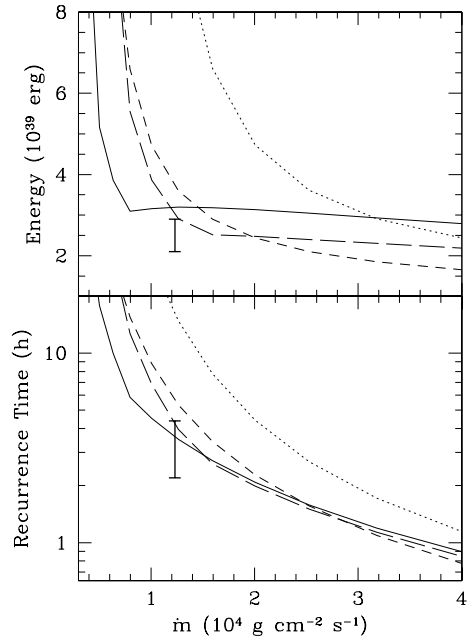}
\vspace{-0.6 cm}
\caption{Predicted burst energy and recurrence time as a function of
assumed local accretion rate $\dot m$ for 4U~1820-30
(ref.\cite{C03}). Choosing $\dot m$ allows both pure helium models
(dotted and short-dashed lines) and models with hydrogen (long-dashed
line 10\% hydrogen; solid line 20\% hydrogen) to fit the observed
values (indicated by error bars and placed at the value of $\dot m$
inferred from the X-ray luminosity).  }
\label{fig:1820}
\end{figure}

The extremely short orbital period implies a hydrogen-deficient
companion, and observations of bursts support this picture, with
$\alpha\approx 120$ \cite{Haberl87} as expected for helium rich
material. Different evolutionary models predict that the accreted
matter is either pure helium
\cite{Verbunt87,BailynGrindlay87,Rasio00}, or contains a small amount
of hydrogen (perhaps 5--35\% by mass;
\cite{Tutukov87,FedorovaErgma89,Podsi02}). Recently, I looked at the
question of whether the composition of the accreted material
(specifically the hydrogen fraction) might be inferred from the burst
properties \cite{C03}. This is important to test intermediate-mass
binary evolution models for X-ray binaries (e.g.~\cite{Podsi02}) which
predict a small amount of hydrogen may be present.

The inferred accretion rate when bursts are seen from 4U~1820-30 is
$\dot M\approx 2.4\times 10^{-9}\ M_\odot\ {\rm yr^{-1}}$, giving an
ignition column depth $\approx 2\times 10^8\ {\rm g\
  cm^{-2}}$. Bildsten \cite{Bildsten95} found that this compares well
with time-dependent models of pure helium burning, although for a
slightly hotter base temperature than expected. The reason for this
can be seen by inspecting Figure \ref{fig:regimes}. Ignition at
$y\approx 10^8\ {\rm g\ cm^{-2}}$ requires heating equivalent to hot
CNO burning with $Z\sim 0.01$. In the pure helium case, this heating
must be provided by a heat flux into the layer from below, and
translates to a flux equivalent to $\approx 0.4$ MeV per accreted
nucleon, higher than the expected crust luminosity \cite{Brown00}.

If hydrogen is present with mass fraction $X\sim 0.1$, hot CNO burning
during accumulation could provide the required heating \cite{C03}
(this requires $Z\gtrsim 3\times 10^{-3}$, likely satisfied for
4U~1820-30). This is illustrated in Figure \ref{fig:1820}, which shows
the predicted recurrence times and burst energies for models with and
without hydrogen, compared to observed values. Models with hydrogen do
much better for the accretion rate inferred from the X-ray
luminosity. However, it is not possible to reach a firm conclusion
because increasing the accretion rate by only a factor of two allows
pure helium models to come into agreement.

As for GS~1826-24, an additional constraint is required to determine
the accreted composition. One possibility is to look for variations of
the burst energetics with accretion rate. For 10\% accreted hydrogen,
the time to burn all the hydrogen is $\approx 3$ hours, comparable to
the recurrence time. This leads to $\approx 10\%$ variations in burst
fluence as recurrence time varies. This is a promising way to
constrain the accreted composition, but will require further
observations of sequences of regular bursting with different
recurrence times. Another possible constraint comes from simultaneous
modelling of Type I bursts and superbursts, I discuss this further in
\S \ref{sec:interact}. It would also be interesting to compare burst
lightcurves with time-dependent simulations.

\subsection{Global bursting behaviour}

The properties of regular burst sequences from both GS~1826-24 and
4U~1820-30 agree very well with theoretical expectations, and we have
discussed to what extent we can learn about the accreted composition
in these cases. However, it has been pointed out many times that the
global bursting behavior is not understood, and for many bursters is
{\it opposite} to that predicted by theory
\cite{vpp88,Bildsten00}. EXOSAT observations of several Atoll sources
showed that as X-ray luminosity increased, burst properties changed
from regular, frequent bursts ($t_{\rm recur}\approx $ hours) with
long durations ($\approx 30\ {\rm s}$) and low $\alpha\approx 40$, to
irregular, infrequent ($t_{\rm recur}>$ 1 day), short ($<10\ {\rm s}$)
bursts with high $\alpha>100$ and often showing photospheric radius
expansion (PRE) (e.g. \cite{vpp88}). RXTE and BeppoSAX observations
have confirmed this result, with particularly good coverage for the
transient source KS~1731-260 \cite{Muno00,Cornelisse03}. Cornelisse et
al.~\cite{Cornelisse03} found that observations of nine bursters were
consistent with this pattern of bursting, with a universal transition
luminosity $L_X\approx 2\times 10^{37}\ {\rm erg\ s^{-1}}$ (although
not all showed the transition from one burst type to another).

The short bursts are naturally interpreted as pure He bursts; the long
bursts as mixed H/He bursts\footnote{Note that classifying short
  bursts as ``helium'' and long bursts as ``hydrogen'' is appropriate
  for bursts with similar ignition column depths, but care should be
  taken when looking at bursts from a wide range of accretion
  rates. For example, at low accretion rates, the opposite behavior
  may be true, with long, energetic pure helium bursts and short
  hydrogen-triggered bursts. The best quantity to distinguish the
  composition is $\alpha$. The lightcurves of a long mixed H/He burst
  and long pure He burst will also look very different; e.g.~the
  latter will most likely show a large photospheric radius expansion
  episode.}. However, theory predicts a transition from pure He bursts
to mixed H/He bursts as accretion rate increases rather than the other
way around (\S 2). Several explanations have been proposed, all of
which require further investigation.  In the context of 1D models,
mixing of fuel by Rayleigh-Taylor \cite{WallaceWoosley84} or shear
instabilities \cite{Fujimoto87} might allow more rapid burning of
hydrogen; Woosley et al. \cite{Woos03} have again emphasised the
importance of ``compositional inertia'', that bursts ignite on the
ashes of previous bursts; Narayan \& Heyl \cite{NarayanHeyl03} calculate linear modes of
steady burning models, and find fuel is burned before the runaway. Bildsten has stressed that solving this puzzle
may require going beyond spherical symmetry, understanding both the
burning front propagation \cite{Bildsten95}, and the fuel covering
fraction \cite{Bildsten98}.

Some relevant observational facts are (i) regular bursting seen in
GS~1826 is well-understood as mixed H/He bursts, arguing that other
bursters such as KS~1731-260 are in this burning regime when regular
bursting is seen --- if so (unless changes in covering fraction
reverse the local accretion rate trend \cite{Bildsten98}), as $\dot M$
drops to the point where the recurrence time is $t_H$ or longer, a
transition at low $\dot M$ into pure helium bursts should be observed;
(ii) large values $\alpha\sim 1000$ are often seen at high $\dot M$,
implying that both H and He are burning outside bursts; (iii)
4U~1820-30 has a similar luminosity at which regular bursting stops
--- if the same mechanism applies, it must be composition-independent.

Other sources show different behavior. In particular, SAS-3
\cite{Basinska84} and later RXTE \cite{Franco01,vanStraaten01}
observations show that bursts from 4U~1728-34 have helium like
characteristics ($\alpha\approx 110$ and durations $\sim 10\ {\rm
  s}$), but become less energetic, stop showing PRE, and occur more
often as accretion rate increases. The burst fluence is proportional
to the recurrence time; very different to the $\approx$ constant
fluence expected for mixed H/He burning. Detailed models of the bursts
from this source have not been made, but the observations perhaps
suggest accretion of He-rich material, with very little H burning
between bursts.

The difference in bursting behavior is correlated with the spin
frequency measured during Type I bursts. KS~1731-260 \cite{Muno00} has
$\nu_{\rm spin}=524\ {\rm Hz}$; 4U~1728-34
\cite{Franco01,vanStraaten01} has $\nu_{\rm spin}=363\ {\rm Hz}$. For
both sources, burst oscillations are seen in the high accretion rate
state only: for KS1731-260, this is mostly for PRE bursts; for 4U~1728-34,
this is mostly for bursts without PRE. Muno and coworkers have shown that the
correlation of the presence of burst oscillations with or without PRE
applies to other sources with spins $\approx 600\ {\rm Hz}$ or
$\approx 300\ {\rm Hz}$ respectively \cite{Muno01}. This suggests that
a detailed analysis of burst properties would show that bursters with
$\approx 300\ {\rm Hz}$ spins are similar to 4U~1728-34, and those
with $\approx 600\ {\rm Hz}$ are similar to KS~1731-260. It is not
clear what this is telling us, perhaps that the two groups of systems
have different evolutionary histories, and hence accreted compositions
and burst properties.

\section{SUPERBURSTS}

\subsection{Superbursts as carbon flashes}

A summary of observations of superbursts is given by Kuulkers, this
volume (see also refs.~\cite{Kuulkers02}, \cite{StrohBild03}). Basic
energetic arguments point to ignition of a layer of fuel much more
massive than a typical Type I burst. For a nuclear energy release of 1~MeV per nucleon, $\approx 10^{24}\ {\rm g}$ is needed to match the observed superburst energy of $\approx 10^{42}\ {\rm ergs}$. At an accretion rate $\approx 0.1\ \dot M_{\rm Edd}$ appropriate for these sources, this mass is accreted in a few months. Since helium ignites at much lower masses at these $\dot M$'s (and indeed these sources show ordinary Type I bursts), the superbursts require a new fuel. Theoretical attention has
focused on carbon, which calculations show is present in
the ashes of H/He burning, e.g.~\cite{BrownBildsten98,Schatz03}.

Carbon ignition on accreting neutron stars was studied by several
authors in the 1970's \cite{WoosleyTaam76,TaamPicklum78,LambLamb78}
(originally as a gamma-ray burst model), and more recently for rapidly
accreting neutron stars ($\dot M>\dot M_{\rm Edd}$) by Brown and
Bildsten \cite{BrownBildsten98}. These calculations found ignition
masses of $\sim 10^{26}\ {\rm g}$, giving characteristic energies
$10^{43}$--$10^{44}\ {\rm ergs}$, and recurrence times 10--100 years
at $0.1\ \dot M_{\rm Edd}$. Therefore carbon flashes initially
appeared too energetic to match the $\sim 10^{42}\ {\rm erg}$
superbursts and probably too rare to match the number discovered (the
superburst recurrence time is not well-constrained, but was at most
$\approx 6$ years in one source \cite{Wijnands01}).

However, two pieces of physics act to reduce the observed energies and
recurrence times. Strohmayer \& Brown \cite{StrohBrown} modeled the
superburst in 4U~1820-30 as ignition of a $\approx 10^{26}\ {\rm g}$
layer with $30$\% carbon and 70\% iron by mass. Because this source
accretes helium-rich matter, stable burning is able to produce large
amounts of carbon \cite{BrownBildsten98}. At the peak temperature of
the flash ($>5\times 10^9\ {\rm K}$), neutrino emission is very
efficient and Strohmayer \& Brown found that $\approx 90$\% of the
energy was lost as neutrinos, leaving $\approx 10^{42}\ {\rm erg}$ to
be radiated from the surface in the first few hours.

\begin{figure}[htb]
%\vspace{9pt}
%\framebox[55mm]{\rule[-21mm]{0mm}{43mm}}
%\includegraphics[scale=0.6]{hotcnob.ps}
\includegraphics[scale=0.43]{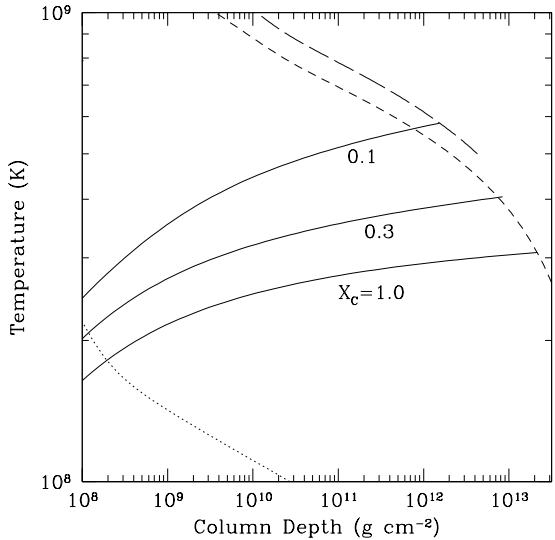}
\vspace{-1.1 cm}
\caption{Temperature profiles in the carbon/heavy element layer (solid
lines) for different carbon mass fractions. The short and long dashed lines show
carbon ignition curves for $X_C=1.0$ and $0.1$; the dotted line shows
the pure helium triple alpha ignition curve for comparison.}
\label{fig:carbign}
\end{figure}

The other superburst sources accrete hydrogen and helium, in which
case protons readily capture on carbon during hydrogen/helium
burning. The amount of carbon produced is small, $\sim 10$\% for
stable burning, or $\lesssim 1$\% for unstable burning
\cite{Schatz03}. However, Cumming \& Bildsten \cite{CB01} showed that
even a small amount of carbon is enough to power a superburst. The key
physics is that the heavy elements that make up most of the mixture
have a low thermal conductivity (this is especially true for the very
heavy nuclei made in the rp-process). This makes a steeper temperature
gradient, leading to ignition at a factor $\approx 10$ lower in column
depth. This is illustrated in Figure \ref{fig:carbign}, which shows
temperature profiles for different carbon mass fractions\footnote{I
  take $\dot M=0.1\ \dot M_{\rm Edd}$, and assume the heating of the
  layer is by a flux from below of 0.1 MeV per nucleon. Brown
  \cite{Browntalk} has recently emphasised that the flux from below is
  very sensitive to the thermal conductivity and superfluid properties
  of the crust \cite{Brown00}, so that (unless the carbon layer has an
  internal heat source) this may be a promising way to constrain the
  thermal state of the interior.}. For $X_C\approx 0.1$, almost all of
the $\approx 10^{42}\ {\rm erg}$ released leaves from the surface,
since neutrino emission is not important at the lower peak
temperature. The recurrence time is $\approx 3\ {\rm years}$ for $y\approx 10^{12}\ {\rm g\ cm^{-2}}$.

\begin{figure}[htb]
%\vspace{9pt}
%\framebox[55mm]{\rule[-21mm]{0mm}{43mm}}
%\includegraphics[scale=0.6]{hotcnob.ps}
\includegraphics[scale=0.43]{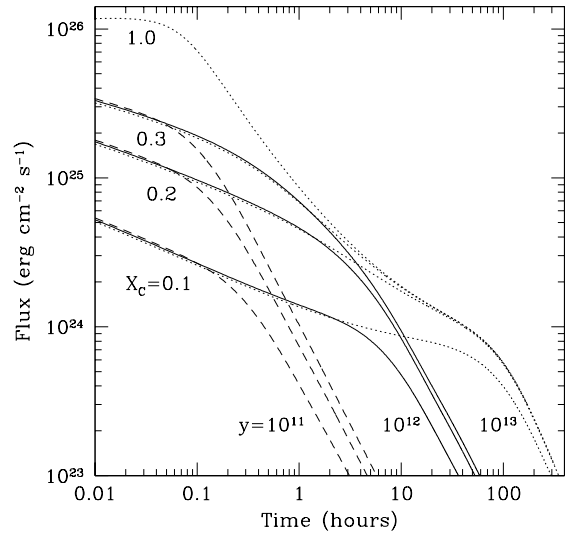}
\vspace{-1.1 cm}
\caption{Surface flux against time for cooling models for
superbursts (Cumming \& Macbeth, in preparation). $X_C$ is the carbon mass fraction, $y$ is the layer column depth (${\rm g\ cm^{-2}}$).}
\label{fig:gop3}
\end{figure}

Superburst sources have a small range of $\dot M$, $\approx 0.1$--$0.25\ \dot M_{\rm Edd}$ (Kuulkers, this volume). Cumming \& Bildsten \cite{CB01} showed that for $\dot
M\lesssim 0.1\ \dot M_{\rm Edd}$, the carbon burns stably before
reaching the ignition curve, explaining the lack of superbursts at low
rates. Another possibility is that this is related directly to the
change in Type I burst behavior observed at $L_X\approx 2\times
10^{37}\ {\rm erg\ s^{-1}}$ in bursters such as KS~1731-260 (\S 3.3;
\cite{Cornelisse03}). Below the transition, where the accreted fuel
burns in regular Type I bursts, the carbon yield is likely very low;
above, where the bursts are short and irregular and not burning all
the fuel, stable burning could produce significant amounts of
carbon. An observational test is to compare bursts
in sources with and without superbursts. The regular burster GS~1826-24 should not show a superburst in this picture. For $\dot M\gtrsim 0.3\ \dot M_{\rm Edd}$, carbon flashes should still occur, but will be less energetic, and show less contrast with the accretion luminosity \cite{CB01}. Long duration $\approx 10^{41}\ {\rm erg}$ bursts are seen from the high $\dot M$ source GX 17+2 \cite{KuulkersGX17}, but may be too frequent to be carbon flashes \cite{CB01}.

\subsection{Time-evolution of the superburst}

A simple model of the superburst lightcurve is to assume the carbon
burns instantaneously, without significant vertical mixing. This may
be a reasonable assumption since the burning occurs much more rapidly
than the convective turnover time (opposite to normal Type I
bursts). The thermal evolution of the layer can then be followed as it
cools, by evolving the entropy equation in time (Cumming \& Macbeth,
in preparation). Figure \ref{fig:gop3} shows a series of lightcurves
for different layer thickness (column depth $y$) and carbon fractions
$X_C$\footnote{In reality, $y$ and $X_C$ are related by the ignition
  conditions, but for simplicity we treat $y$ and $X_C$ as free
  parameters --- this also gives a model independent approach in which
  $X_C$ is just a measure of the deposited energy. For a similar
  approach see \cite{EichlerCheng89}.}.

At early times, as the outer parts of the layer thermally adjust, the
radiative flux depends mostly on carbon fraction. At late times, after
the cooling wave reaches the base of the layer, the flux falls off
more steeply. A toy analytic model of the late time cooling is to take
a slab with constant thermal diffusivity $\kappa$, and perturb the
temperature close to the surface, for example at a depth $a$. For a
delta-function perturbation initially, the temperature evolution is
given by the Green's function for this case, and the surface flux is
$F\propto \exp(-\tau/t)/t^{3/2}$, where $\tau$ is the thermal time at
the initial heating depth $\tau\approx a^2/\kappa$. At late times, the
flux decays as a power law $\propto 1/t^{3/2}$, close to the value
found numerically.

For $y$ in the range $10^{12}$--$10^{13}\ {\rm g\ cm^{-2}}$ and
$X_C=0.1$--$0.3$, we find that $\sim 10^{42}\ {\rm ergs}$ is radiated
from the surface in the first several hours. The observed lightcurves
agree well with these cooling models; a more detailed comparison is in
progress to see whether $X_C$ and $y$ can be constrained.

The calculations shown in Figure \ref{fig:gop3} assume that carbon
burns to iron group after ignition, giving an energy release of
$\approx 1$ MeV per nucleon. However, Schatz et al.~\cite{SB03} showed
that additional energy is released by photodisintegration of elements
heavier than iron into iron group nuclei. The binding energy released
in going from these heavy nuclei into iron is $\approx 0.1$ MeV per
nucleon, ten times less than carbon burning into iron, but for
$X_C\lesssim 10$\%, this contribution dominates the energetics. This
is an interesting additional way in which the nuclei made in the rp
process directly affect the superburst energetics.

\subsection{Interaction between normal Type I bursts and superbursts}
\label{sec:interact}

Type I bursts disappear (are ``quenched'') for $\approx $ weeks
following the superburst. Cumming \& Bildsten \cite{CB01} proposed
that the cooling flux from the superburst stabilizes the hydrogen and
helium burning. Assuming an exponential cooling law, they estimated a
quenching time of $\approx 5 t_{\rm cool}\approx $ days, where $t_{\rm
  cool}$ is the cooling time for the layer. This is a little shorter
than the observed timescales of $\approx $ weeks (Kuulkers, this
volume). However, the cooling models in Figure \ref{fig:gop3} show a
slow power law decay, $F\propto t^{-3/2}$, at late times rather than
exponential. Inspection of Figure \ref{fig:gop3} shows that the flux
reaches $\approx 10^{24}\ {\rm erg\ cm^{-2}\ s^{-1}}$ in a time
$\approx 10\ {\rm h}\ (y/10^{12}\ {\rm g\ cm^{-2}})$, followed by a
$t^{-3/2}$ decay. Assuming the stabilizing flux is $\approx 10^{22}\
{\rm erg\ cm^{-2}\ s^{-1}}$ \cite{Paczynski83,Bildsten95} gives a
quenching time
\begin{equation}
t_{\rm quench}\approx 22\ t_{\rm cool}\approx 9\ {\rm days}\
\left({y\over 10^{12}\ {\rm g\ cm^{-2}}}\right),
\end{equation}
consistent with observed timescales. Turning this around, a
measurement of the quenching time allows the thickness of the layer to
be determined.

Whenever the rise of the superburst has been seen, there is a
precursor event that resembles a normal Type I burst. A likely
explanation for this is that flux from deep carbon burning heats and
ignites the layer of H/He accumulating on the surface. This scenario
has yet to be investigated in detail, and in particular outstanding
questions are the relative timing between precursor and superburst,
and whether the vertical propagation of the burning during the rise
triggers the precursor. However, simple predictions of this model are:
(i) most superbursts should have a precursor (but not all, since a
small fraction of the time the helium layer will be too thin to
unstably ignite), (ii) a thicker helium layer will ignite with a lower
flux, perhaps giving a correlation between precursor energy and the
time until the superburst rise, and (iii) the precursor should be
weaker than Type I bursts at the same $\dot M$, since the mass of the
H/He layer is less than the critical mass.

The final point is that simultaneous modelling of normal Type I bursts
and superbursts gives an important additional constraint on the models
of both \cite{C03}. For example, Strohmayer and Brown found a recurrence time of 13 years in their models of the superburst from 4U~1820-30 \cite{StrohBrown}, in which they inferred the accretion rate from the X-ray luminosity. However, using the regular Type I bursting to calibrate the accretion rate gives a shorter recurrence time, 1--2 years for pure
helium, 5--10 years for 10\% hydrogen \cite{C03}.

\section{CONCLUSIONS}

The basic theory of nuclear burning on accreting neutron stars (first
outlined over twenty years ago, e.g.~\cite{FHM}) has mixed success in
explaining observed Type I X-ray burst properties. The extremely
regular burster GS~1826-24 shows remarkable agreement with models of
mixed H/He bursts. The observed variation of $\alpha$ with $\dot M$
strongly suggests that the accreted material has solar metallicity
\cite{Gall03}. Regular burst sequences from 4U~1820-30 agree well with
accretion and burning of helium-rich matter. For this system,
observations of burst fluence variations with recurrence time would
constrain the amount of hydrogen present in the accreted material
\cite{C03}, relevant for testing evolutionary models.

Superbursts fit naturally into the picture as ignition of fuel at a
much larger depth than normal Type I bursts. Calculations of H/He
burning show that the heavy ashes contain a small amount of carbon
($\sim 1$--$10$\%). Ignition calculations show that this mixture will
unstably ignite giving rise to a burst with properties matching
observed properties of superbursts. Time dependent models of the
superburst lightcurve show good agreement with the observed
exponential decay times and energies. Because stable burning produces
more carbon than unstable burning \cite{Schatz03}, a self-consistent
model of carbon production and ignition may require understanding the
nature of H/He burning at high accretion rates where bursts are
irregular and fuel burns between bursts. There is much to be learned
from the interactions between superbursts and normal Type I
bursts. The properties of precursors and the time of cessation of
bursting after a superburst will tell us much about the underlying
layers.

The global Type I bursting behavior remains the biggest puzzle, and
may well need new physics to explain it. Another intriguing problem is
how to explain bursts with ten minute recurrence times (e.g.~see
\cite{Gottwald86}), not predicted by theory. There are many
observational constraints which will allow progress to be made:
variations of burst properties with accretion rate, shape of burst
lightcurves, interactions between normal Type I bursts and
superbursts, and the properties of burst oscillations. We are slowly
learning more about bursts at higher or lower luminosities than
traditional bursters
(e.g.~\cite{Cornelisse02,KuulkersGX17}). Understanding the rich
phenomenology of bursts promises to teach us much about these neutron
stars and the physical processes in their outer layers.

I thank L.~Bildsten, R.~Cornelisse, D.~Galloway, E.~Kuulkers, and J.~Macbeth for comments on the manuscript. D.~Galloway and A.~Heger kindly provided data for Fig.~3. I acknowledge support from NASA through Hubble Fellowship grant HF-01138 awarded by the Space Telescope Science Institute, which is operated by the
Association of Universities for Research in Astronomy, Inc., for NASA,
under contract NAS 5-26555.

\end{document}